# RHR-Net: A Residual Hourglass Recurrent Neural Network for Speech Enhancement

*Jalal Abdulbaqi, Yue Gu, and Ivan Marsic*

Rutgers, the State University of New Jersey, USA

`jalal.nazar@rutgers.edu, yue.guapp@rutgers.edu, marsic@rutgers.edu`

## Abstract

Most current speech enhancement models use spectrogram features that require an expensive transformation and result in phase information loss. Previous work has overcome these issues by using convolutional networks to learn long-range temporal correlations across high-resolution waveforms. These models, however, are limited by memory-intensive dilated convolution and aliasing artifacts from upsampling. We introduce an end-to-end fully-recurrent hourglass-shaped neural network architecture with residual connections for waveform-based single-channel speech enhancement. Our model can efficiently capture long-range temporal dependencies by reducing the features resolution without information loss. Experimental results show that our model outperforms state-of-the-art approaches in six evaluation metrics[1].

**Index Terms**: speech enhancement, speech denoising, recurrent neural network, raw waveform, residual connection

## 1. Introduction

Speech enhancement has important applications in voice communication, hearing aids, and automatic speech recognition. Speech enhancement removes background noise from noisy speech signals, increasing speech quality and intelligibility [1], [2]. Early research used non-trainable statistical approaches on spectrograms, such as spectral subtraction [3], Wiener filter [4], statistical model-based methods [5], the subspace method [6], minimum mean-square error estimator, and optimally-modified log-spectral amplitude [7], [8]. These methods showed limited performance on speech with non-stationary noise, which is common in real-life environments. Non-negative matrix factorization has later been widely used for speech separation and enhancement [9], [10].

Recently, deep neural networks have been employed to overcome the non-stationary condition and have improved speech quality and intelligibility. Early models used mapping-based methods, where the enhanced signal is directly predicted from the noisy one. Several such deep learning models have been developed, including denoising autoencoders [11] (using fully-connected layers), recurrent neural networks (RNN) [12] and convolutional neural networks (CNN). Later, a masking-based method was introduced to enhance the signal by applying the noisy signal to the predicted mask [13]–[17].

Most of these methods use time-frequency (T-F) spectrogram features instead of time-domain waveform, since T-F has a reduced resolution. Spectrogram features, however, have certain limitations. First, the pre- and post-processing operations such as discrete Fourier transform and its inverse are computationally expensive, and cause artifacts in the output signal. Second, these approaches usually only estimate the magnitude, and use the noisy phase to produce the enhanced speech. Research has shown that the phase can enhance the speech quality [18]. Recent research has considered predicting the phase and the magnitude at the cost of model complexity, such as adding special model for phase component [19], or implementing a complex-valued neural network [13].

Recently, several studies proposed overcoming previous limitations by working directly on the waveform. Fu et al. [20] compared fully-convolutional networks with fully-connected networks. Pascual et al. [21] implement a generative adversarial network for speech enhancement (SEGAN), using strided convolutions, residual connections, and an encoder-decoder architecture. Later, a text-to-speech model called Wavenet [22] directly synthesized raw waveforms. Qian et al. [23] and Rethage et al. [24] presented a modified version of WaveNet for speech denoising. The former integrated a Bayesian framework WaveNet, while the latter used a non-causal dilated convolution with residual connections. Germain et al. [25] presented dilated convolutions combined with a feature loss network. Stroller et al. [26] adapted the U-Net [27] model for source separation using dilated convolutions and linear interpolation instead of transposed convolution for upsampling. All these methods used convolutional neural networks due to their ability to capture the samples' dependencies better than fully-connected networks. Because waveform is a sequential datatype, it requires a temporal context as well. Recurrent neural networks are known to capture temporal sequence information [28] and are used in many sequential applications such as speech recognition, neural machine translation, and spectrogram-based speech enhancement. However, to our knowledge only [29] has applied RNN for denoising a non-speech waveform and no one has used it for waveform-based speech enhancement. The reason may be that the high resolution of waveforms requires more expensive, deeper, and wider networks. It is difficult to build a deep RNN because of saturating activation function, which causes gradient decay over layers. Also, we found empirically that RNNs sufficiently wide to process the high-resolution waveforms exceeded the available memory capacity. We, therefore, introduce our residual hourglass recurrent neural network (RHR-Net) for waveform-based single-channel speech enhancement. RHR-Net overcomes the RNN limitations by introducing two techniques. First, the network architecture has an hourglass shape; the layers in the lower pyramid reduce the number of time-steps and increases the number of units (width), while the upper pyramid does the reverse. This architecture allows the RNN to handle high-resolution waveform features without memory overflow. Second, using residual connections between the same-shaped layers from the lower pyramid to the upper one prevents gradient decay over layers and improves the model generalization.

---

[1] Audio samples are available at the following link:
https://jalal-abdulbaqi.github.io/AudioSamples/

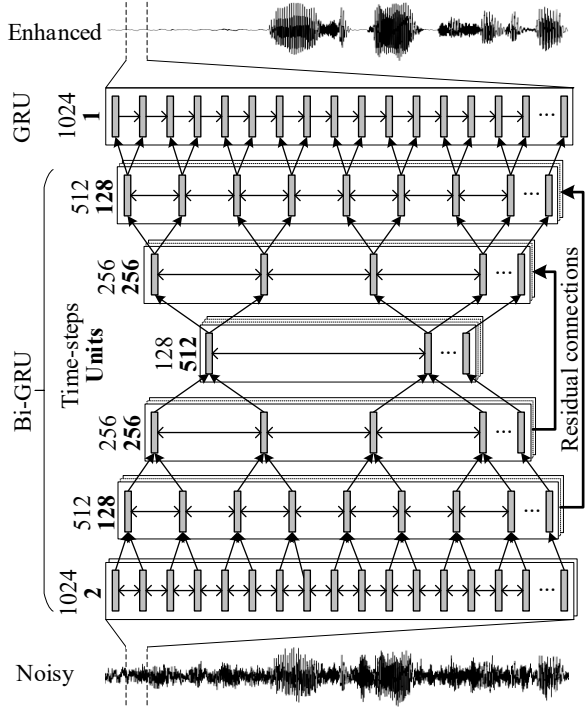

Figure 1: *Our proposed RNN architecture. Seven stacked RNN layers with the numbers on the left representing the number of time steps and the number of units in each layer. Wider layers have fewer units and vice versa. The two bold arrows on the right represent the residual connections.*

Advantages of our model:

- Uses a raw waveform, without any transformation or handcrafted features.
- Does not lose information at upsampling layers, unlike linear interpolation methods.
- Is a simple end-to-end design that outperforms several more complex neural network approaches.
- Is a novel deep RNN architecture that can be applied for regression problems other than speech enhancement.

We evaluated our model using six objective metrics, demonstrating its ability to significantly enhance speech quality and intelligibly. The next section provides an overview of the model architecture. Section 3 describes the dataset we used and the preprocessing operations. Section 4 presents the experimental setup and discusses the results. Section 5 concludes and suggests future work.

## 2. Model Architecture

Our model includes seven GRU layers with two residual connections. The first six layers are bidirectional and the last one is a single GRU (Figure 1). The goal of our speech enhancement network is to learn non-linear relationships, so that noisy speech $y(t)$ can be translated into clean speech $x(t)$:

$$y(t) = f(x(t)) \quad (1)$$

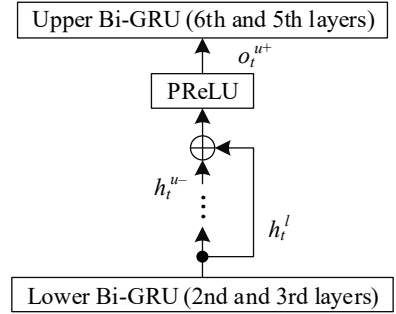

Figure 2: *A high-level view highlighting the residual connections in our proposed model from Figure 1.*

The input vector $X = (x_1, \ldots, x_T)$ represents a $T$-seconds wide segment from a noisy audio waveform signal.

RNNs can efficiently realize temporal features in sequential data, so they have been used widely to process speech data either for speech recognition or enhancement. We chose gated recurrent units (GRU) instead of long short-term memory units (LSTM) or vanilla RNN. Both GRU and LSTM outperform vanilla RNN [28], but GRUs have a simpler structure and train faster than LSTMs. In addition, we chose bidirectional RNNs since in speech enhancement each predicted sample can depend on future as well as past noisy samples. The stacked GRU increases the capacity of the network by sharing the hidden states not only from the same layer but also from the lower layers as well. The stacked bidirectional RNNs share their hidden states, so that the hidden state ($h_t^l$) of a bi-GRU unit in layer $l$ at time $t$ is obtained by concatenating its forward ($\overrightarrow{h_t^l}$) and backward ($\overleftarrow{h_t^l}$) hidden states, which depend on the lower layer $l$–1 at time $t$ and this layer at time $t$–1:

$$\overrightarrow{h_t^l} = \overrightarrow{GRU}\left(\overrightarrow{h_t^{l-1}}, \overrightarrow{h_{t-1}^l}\right) \quad (2)$$

$$\overleftarrow{h_t^l} = \overleftarrow{GRU}(\overleftarrow{h_t^{l-1}}, \overleftarrow{h_{t-1}^l}) \quad (3)$$

$$h_t^l = conc(\overrightarrow{h_t^l}, \overleftarrow{h_t^l}) \quad (4)$$

The two pyramids of our hourglass architecture keep the number of trainable parameters within the memory constraints. The bottom pyramid decreases the number time steps while increasing the number of GRU units per layer, and the top pyramid does the reverse. This approach allows for deeper networks. We did not use upsampling techniques, such as linear interpolation, because the information can be lost. Instead, we reshape the RNN output to the desired fewer time steps. Reshaping the layer output to decrease and increase the time steps prevents losing data, and allows the RNN to have a sufficient size of units. However, while stacking RNNs can increase the capacity of the network, deeper RNNs usually have gradient decay issues due to their saturating activation functions. To address this issue, we used residual connections between the lower and upper layers (Figures 1 and 2). The residual connections facilitate training the deep RNN, and provide better generalization by combining the low-level features with the high-level ones in the upper layers. In Figure 2, the hidden states of the lower layer ($h_t^l$) and those of

Table 1: *Evaluation results of our proposed model compared with other state-of-the-art research work using six objective metrics on the same dataset* [30]. *Higher scores are better, and the highest scores are boldfaced.*

| Model | Features type | SSNR | PESQ | STOI | CSIG | CBAK | COVL |
|---|---|---|---|---|---|---|---|
| No Enhancement (Noisy) | - | 1.68 | 1.97 | 0.820 | 3.35 | 2.44 | 2.63 |
| SEGAN, 2017 [21] | waveform | 7.73 | 2.16 | 0.93 | 3.48 | 2.94 | 2.80 |
| CNN-GAN, 2018 [15] | spectrogram | - | 2.34 | 0.93 | 3.55 | 2.95 | 2.92 |
| Wavenet, 2018 [24] | waveform | - | - | - | 3.62 | 3.23 | 2.98 |
| MMSE-GAN, 2018 [17] | spectrogram | - | 2.53 | - | 3.80 | 3.12 | 3.14 |
| DFL, 2018 [26] | waveform | - | - | - | 3.86 | 3.33 | 3.22 |
| Large-DCUnet-20, 2019 [13] | spectrogram | 14.68 | **3.22** | - | 4.33 | 3.96 | 3.79 |
| *RHR-Net (Our model)* | waveform | **14.71** | 3.20 | **0.98** | **4.37** | **4.02** | **3.82** |

the upper layer before the residual connection ($h_t^{u-}$) are combined to produce the residual output:

$$o_t^{u+} = PReLU(h_t^l + h_t^{u-}) \quad (5)$$

where $PReLU$ is the parametric rectified linear unit activation function. Finally, we use a single forward GRU to output the enhanced speech with the same size of the input vector:

$$\overrightarrow{h_t^l} = \overrightarrow{GRU}\left(\overrightarrow{h_t^{l-1}}, \overrightarrow{h_{t-1}^l}\right) \quad (6)$$

Therefore, the output will be created by combining the hidden states for each input segment:

$$Y = (\overrightarrow{h_1^7}, ..., \overrightarrow{h_T^7}) \quad (7)$$

where $Y$ denotes the enhanced signal output and $\overrightarrow{h_1^7}$ denotes the hidden state of the last (seventh) layer.

## 3. Dataset and Preprocessing

The dataset used for training and evaluating our model has been set up in [30]. We chose this dataset because it is large, has different types of non-stationary noise, and is public so that we can compare our results with other published work. This dataset is an excerpt of the Voice Bank corpus [31] with 28 speakers (14 male and 14 female) of the same accent region (England) and another 56 speakers (28 male and 28 female) of other accent regions (Scotland and United States).

The noisy data used for training are two artificially generated (speech shaped noise and babble) and eight real noise recordings from the Demand database [32]. The noises are from different environments such as kitchens, offices, public spaces, transportation stations, and streets. The training set includes 11,572 utterances with five signal-to-noise (SNR) values: 15 dB, 10 dB, 5 dB, and 0 dB.

The noisy data used for testing include two other speakers of the same corpus from England (a male and a female), and five other noises from the Demand database. The chosen noises include a living room, an office, a bus, and street noise. The testing set includes 824 utterances with five SNR: 17.5 dB, 12.5 dB, 7.5 dB and 2.5 dB. We downsampled the audio signals to 16kHz, getting a reasonable dataset size for recognizing speech. Our preprocessing included slicing both noisy and clean speech signals into 1024 samples (~64 ms) with 25% overlap during training and without overlap during the evaluation. We did not use any other preprocessing, such as pre-emphasis.

## 4. Experiment Setup and Results

Our architecture uses seven GRU layers. The first six are bi-directional, while the last one is single-directional to produce the enhanced signal (Figure 1). The number of units per layer are: 2, 128, 256, 512, 256, 128, and 1; the size of the time steps per layer are: 1024, 512, 256, 128, 256, 512, and 1024. Two residual connections link the second and third layers with the sixth and fifth, respectively. The PReLU activation function is used with residual connections, as it does not saturate the negative values compared to Leaky-ReLU and has been shown to improve model fitting [33]. The model has 2 million trainable parameters, which is small with respect to Wavenet which has 6.3 million. We use the Xavier normal initializer [34] for the kernel weights, with zero-initialized biases. Xavier initialization keeps the values of the weights in a reasonable range, preventing the inputs from shrinking or growing more than needed through the layers. It determines the initialization values with respect to the number of input and output neurons. The initializer for the recurrent states is a random orthogonal matrix [35], which helps the RNN stabilize by avoiding vanishing or exploding gradients. The stability occurs because the orthogonal matrix has an absolute eigenvalue of one, which avoids the gradients from exploding or vanishing due to repeated matrix multiplication.

We use the log-cosh loss function, a regression loss function that takes on the behavior of squared-loss when the loss is small, and absolute loss when the loss is large; this reduces the influence of wrong predictions. The optimization algorithm used is RMSprop [36], which helps the training of large neural networks on large redundant datasets. In addition, Keras [37] documentation recommends using this algorithm with RNN. We trained the model until the validation loss increased with a batch size of 512, using two NVIDIA GTX-1080 GPUs. We used different learning rates during training, starting at $10^{-4}$ and gradually decreasing to $10^{-8}$. The library used to implement this work was Keras with TensorFlow [38] as a backend. To evaluate our model, we computed six objective metrics using an open-source implementation[1,2]:

- Segmental signal-to-noise ratio (SSNR) [1]: computed by dividing the clean and enhanced signals into segments and computing the segment energies and SNRs, and then returning the mean segmental SNR (dB). The values range from -10 to 35.

---

[1] https://www.crcpress.com/downloads/K14513/K14513_CD_Files.zip

[2] http://ceestaal.nl/stoi.zip

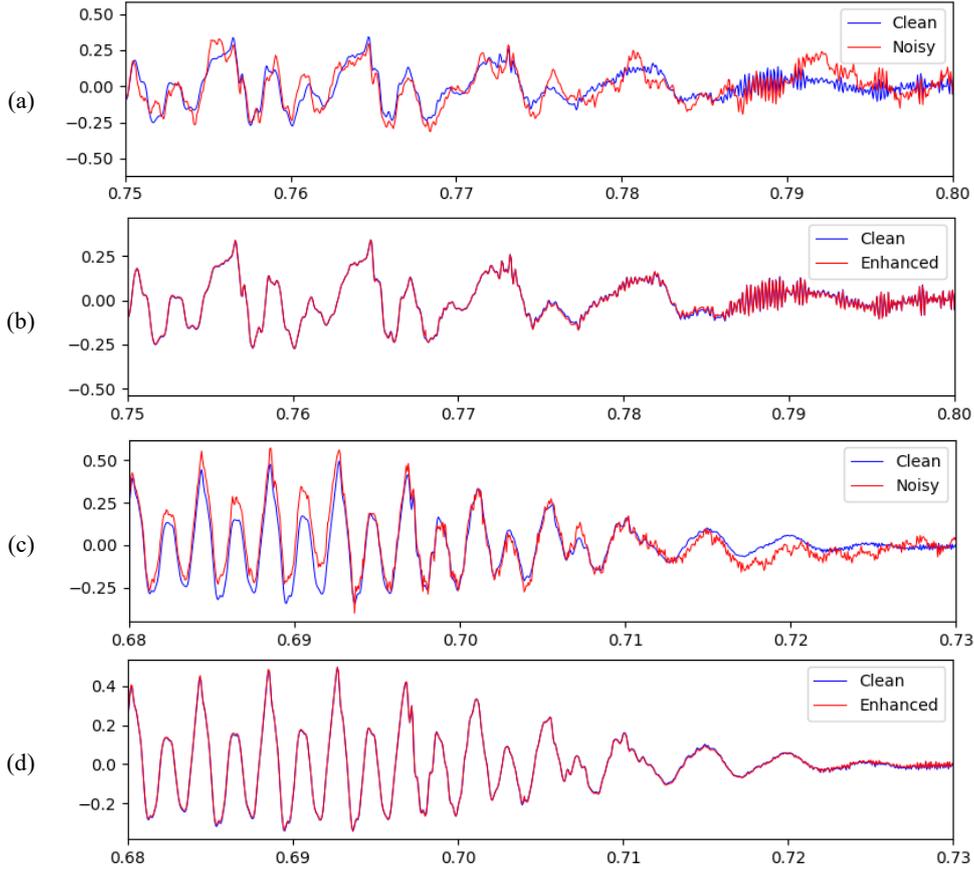

Figure 3: *An illustration of speech enhancement using our model using speech samples with SNR = 2.5 dB and duration of 50 msec from the test dataset.(a) The sample number 232_052. The blue lines represent the clean speech and the red lines represent the noisy speech.(b) The corresponding enhanced speech (red line) compared with the clean input speech (blue line) (c) The speech sample 257_054. (d) The corresponding enhanced speech.*

- Perceptual evaluation of speech quality (PESQ) [1]: a more complex metric to capture a wider range of distortions. PESQ is the most common metric to evaluate the speech quality, calculated by comparing the enhanced and clean speech. The values range -0.5 to 4.5.
- Short-time objective intelligibility (STOI) [39]: reflects the improvement in speech intelligibility with a score range from 0 to 1.
- Three subjective mean opinion scores (MOSs): CSIG for signal distortion evaluation, CBAK for noise distortion evaluation, and COVL for overall quality evaluation. We used their mathematical representations, and their scores range from 1 to 5 [1].

For all these metrics, higher values mean better performance. Two speech test samples (small segment 50 ms) are illustrated in Figure 3. Both samples include non-stationary noise with people talking in background ("cocktail party") and music playing. For each segment, the foreground speaker talks (high frequency) in the first half, while foreground speaker stops talking (low frequency) in the second half. The enhanced speech signal tracks the clean in both cases, which shows the model ability to capture the clean speech in all speaker events.

Our model outperformed all the previous models by a considerable margin (Table 1). Our model achieved better performance than Large-DCUnet-20 [13] in all three MOS scores and segmented SNR, and only barely fell short in PESQ. Our simple and efficient seven-layer RNN model outperformed the Deep Complex U-Net model, which used 20 layers and used a complex-number masking approach for magnitude and phase.

## 5. Conclusion

We introduced a novel end-to-end fully-recurrent neural network for single-channel speech enhancement. Our recurrent layers are designed in a hourglass shape to reduce the speech signal dimension and assist recognition of the long-term dependencies. The results show that our simple and efficient model outperforms most of the current approaches with more complex architectures. We will evaluate this model with other datasets and apply to other sequential applications.

## 6. Acknowledgements

We thank Shuhong Chen for his help. This research was supported by the National Institutes of Health under Award Number R01LM011834.